\documentclass[USenglish,twocolumn]{article}

\usepackage[utf8]{inputenc}
\usepackage[big]{dgruyter}

\begin{document}
 
  \articletype{Research Article{\hfill}Open Access}

  \author*[1]{Spiegel Lubov}

  \author[2]{Polyachenko Evgeny}



  \title{\huge Multiarm spirals on the periphery of disc galaxies}

  \runningtitle{Multiarm spirals}


  \begin{abstract}
{Spiral patterns in some disc galaxies have two arms in the centre, and three or more arms on the periphery. The same result is also obtained in numerical simulations of stellar and gaseous discs. We argue that such patterns may occur due to fast cooling of the gas, resulting in formation of giant molecular clouds. The timescale of this process is 50 Myr, the factor of 10 shorter than of ordinary secular instability. The giant molecular clouds give rise to multiarm spirals through the mechanism of swing amplification.}
\end{abstract}

  \journalname{Open Astronomy}
\DOI{DOI}
  \startpage{1}
  \received{..}
  \revised{..}
  \accepted{..}

  \journalyear{2017}
  \journalvolume{1}

\maketitle


{ \let\thempfn\relax
\footnotetext{\hspace{-1ex}{\Authfont\small \textbf{Corresponding Author: Spiegel Lubov,}} {\Affilfont Institute of Astronomy, Russian Academy of Sciences, 48 Pyatnitskya St., Moscow 119017, Russia; E-mail: spiegel@inasan.ru}}
}

{ \let\thempfn\relax
\footnotetext{\hspace{-1ex}{\Authfont\small \textbf{Polyachenko Evgeny,}} {\Affilfont Institute of Astronomy, Russian Academy of Sciences, 48 Pyatnitskya St., Moscow 119017; E-mail: epolyach@inasan.ru}}
}

\section{Introduction}
\begin{figure*}
  \centerline{
  \includegraphics[width = 40mm]{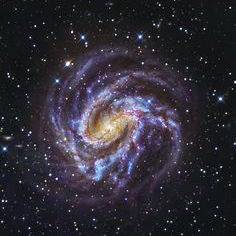}
  \includegraphics[width = 40mm]{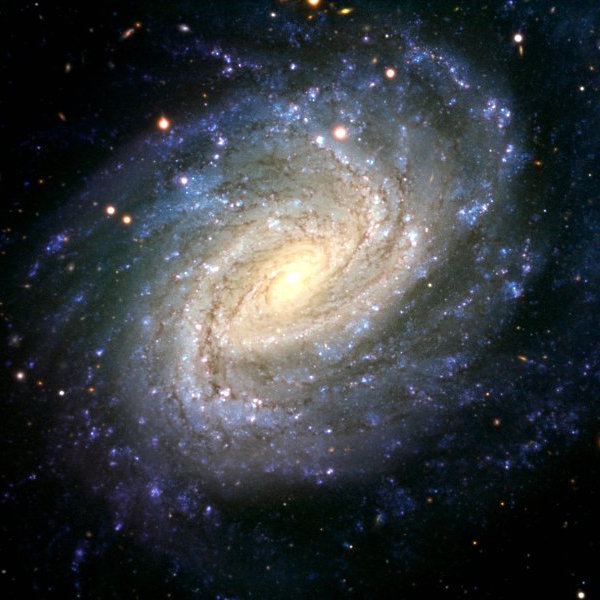}
  \includegraphics[width = 40mm]{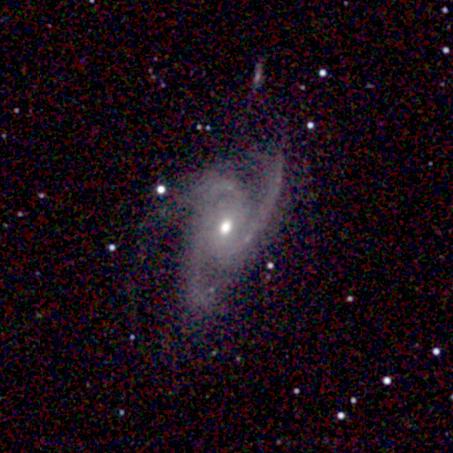}
  \includegraphics[width = 40mm]{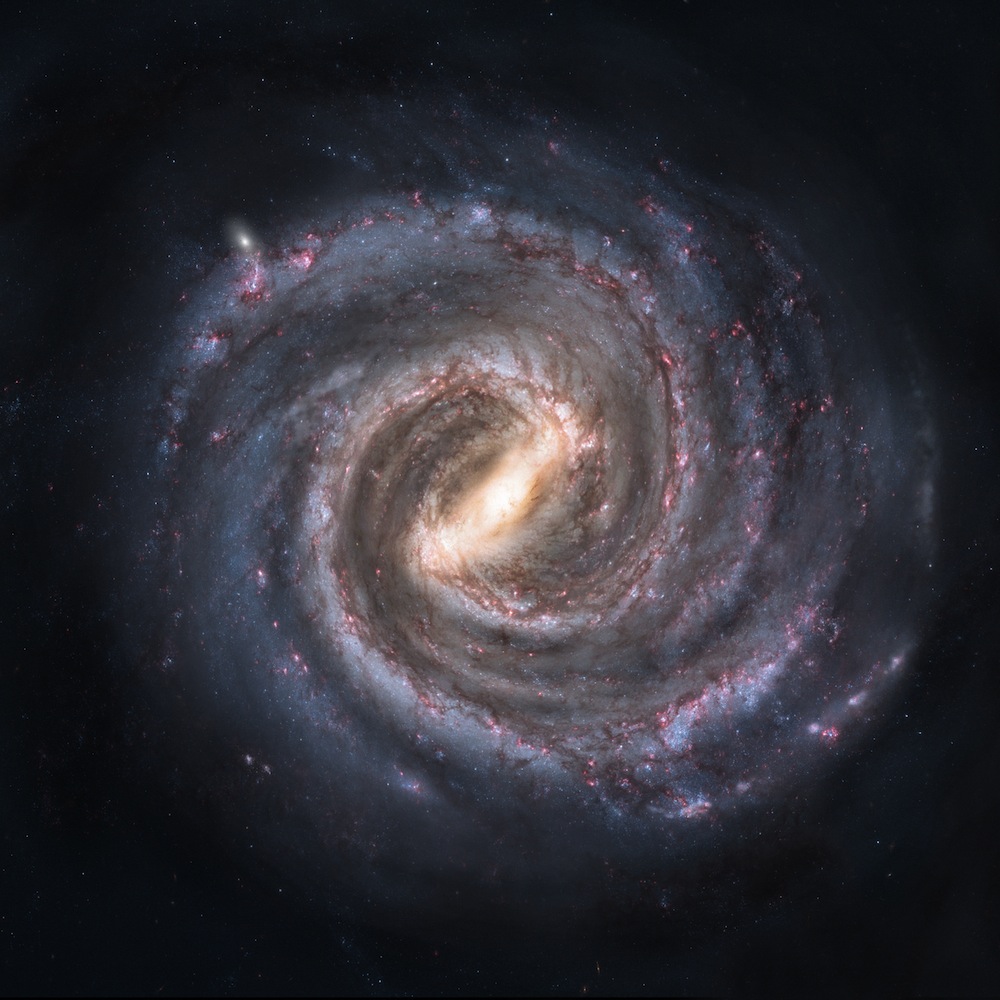}}
  \caption{Examples of disc galaxies with three or more peripheral arms (from the left): M83 (RGB, Anglo-Australian Observatory), NGC 1187 (BV, VLT), NGC 5054 (JHK, 2MASS), `Milky Way' (reconstruction).}
  \label{fig:gal}
\end{figure*}

The spiral pattern of disc galaxies does not always have the form of two, symmetrically located arms extending from the centre or the central bar to the edge of the disc. Often the pattern has two arms in the central region, and more arms in the outer parts (see Fig.\,\ref{fig:gal}). The same phenomenon was observed in recent stellar-gas simulation by \citet[][hereafter K+]{KKK2016}. According to their conclusion, the presence of the gas component radically changes the evolution of the galaxy, despite small mass of the gas.

If gas evolution is neglected, then slow formation of the stellar bar without appreciable spirals is observed, with the e-folding time $T_\textrm{e} \approx 500$\,Myr. In the presence of the active gas component, a two-arm spiral is formed at the centre of the disc and a three-arm spiral is formed at the periphery, with the formation rate being approximately an order of magnitude higher. The purpose of this paper is to analyse the model of the Galaxy used by K+, and to suggest a mechanism for multiple arm formation.

\section{The Galaxy model}

Following K+ we adopt a model consisting of two active and two fixed components. The active components are a thin stellar disc and a gaseous disc that governed by the corresponding dynamical equations. The fixed components (a halo and a bulge) are given by an unchanged external gravitational potential.

For the thin stellar disc, an exponential surface density profile is adopted:
\begin{equation}
	\Sigma_\textrm{0}(R) = \Sigma_{*} \textrm{e}^{-R/R_\textrm{d}}\ ,  
\end{equation}
where $R_\textrm{d} = 3$\,kpc and $M_\textrm{d} = 2\pi R^2_\textrm{d}\Sigma_{*} = 4\cdot 10^{10}$\,M$_\odot$ are the radial scale length and the mass of the disc. A radial velocity dispersion profile is
\begin{equation}
	\sigma_R(R) = \sigma_0 \textrm{e}^{-R/R_\sigma}\ ,  
\end{equation}
with $R_\sigma = 6.4$\,kpc and $\sigma_0 = 120$\,km/s, so that the dispersion in the solar neighbourhood is 34\,km/s.

In our model the density and turbulent sound speed of the gaseous discs are smooth functions of radius: instead of the  gaseous disc of mass $M_\textrm{g} = 4\cdot 10^{9}$\,M$_\odot$ with flat density profile $\Sigma_\textrm{g0}=15.7$\,M$_\odot/$pc$^2$ trimmed at $R=R_\textrm{g} = 9$\,kpc, we use Kuzmin-Toomre disc
\begin{equation}
	\Sigma_\textrm{g0}(R) = \frac{\Sigma_{\textrm{g}*}}{(1+R^2/R^2_\textrm{g})^{3/2}}
\end{equation}
with central surface density $\Sigma_{\textrm{g}*}=39.7\textrm{\,M}_\odot/\textrm{pc}^2$ and the radial scale $R_\textrm{g} = 6$\,kpc.
For these parameters, the mass of the disc inside $R = 9$\,kpc is still equal to $4\cdot 10^{9}$\,M$_\odot$, and the gas density in the solar neighbourhood turns out to be 8.6\,M$_\odot/$pc$^2$. Also, in our work we have adopted a sound speed profile that is different from the constant $c_\textrm{s} = 8$\,km/s adopted in K+:
\begin{equation}
	c_\textrm{s}(R) = \frac{c_0}{(1+R^2/R^2_\textrm{g})^{1/2}}\ ,  
\end{equation}
where $c_0 = 5.8$\,km/s, and $3.5$\,km/s in the solar neighbourhood.

For the halo potential, the pseudo-isothermal profile is selected,
\begin{equation}
	\rho_\textrm{h}(r) = \frac{\rho_\textrm{h0}} {(1 + r^2/a^2_\textrm{h})}\ ,
\end{equation}
where $a_\textrm{h} = 3$\,kpc is the characteristic halo scale, and the central density is $\rho_\textrm{h0} = 0.085\textrm{\,M}_\odot/\textrm{pc}^3$ is determined from the condition that the halo mass inside the radius $r_\textrm{h} = 12$\,kpc is $7.7\cdot10^{10}$\,M$_\odot$.

The bulge is given by the Plummer distribution
\begin{equation}
	\rho_\textrm{b}(R) = \frac{\rho_\textrm{b0}}{(1+R^2/R^2_\textrm{b})^{5/2}}\ ,  \quad \rho_\textrm{b0}=47.6\textrm{\,M}_\odot/\textrm{pc}^3\ ,
\end{equation}
with the bulge scale $R_\textrm{b} = 0.44$\,kpc. This density profile suggests the bulge mass within $r=1.5$\,kpc equal to $1.5\cdot10^{10}$\,M$_\odot$.

\begin{figure}
  \centerline{\includegraphics[width = 85mm]{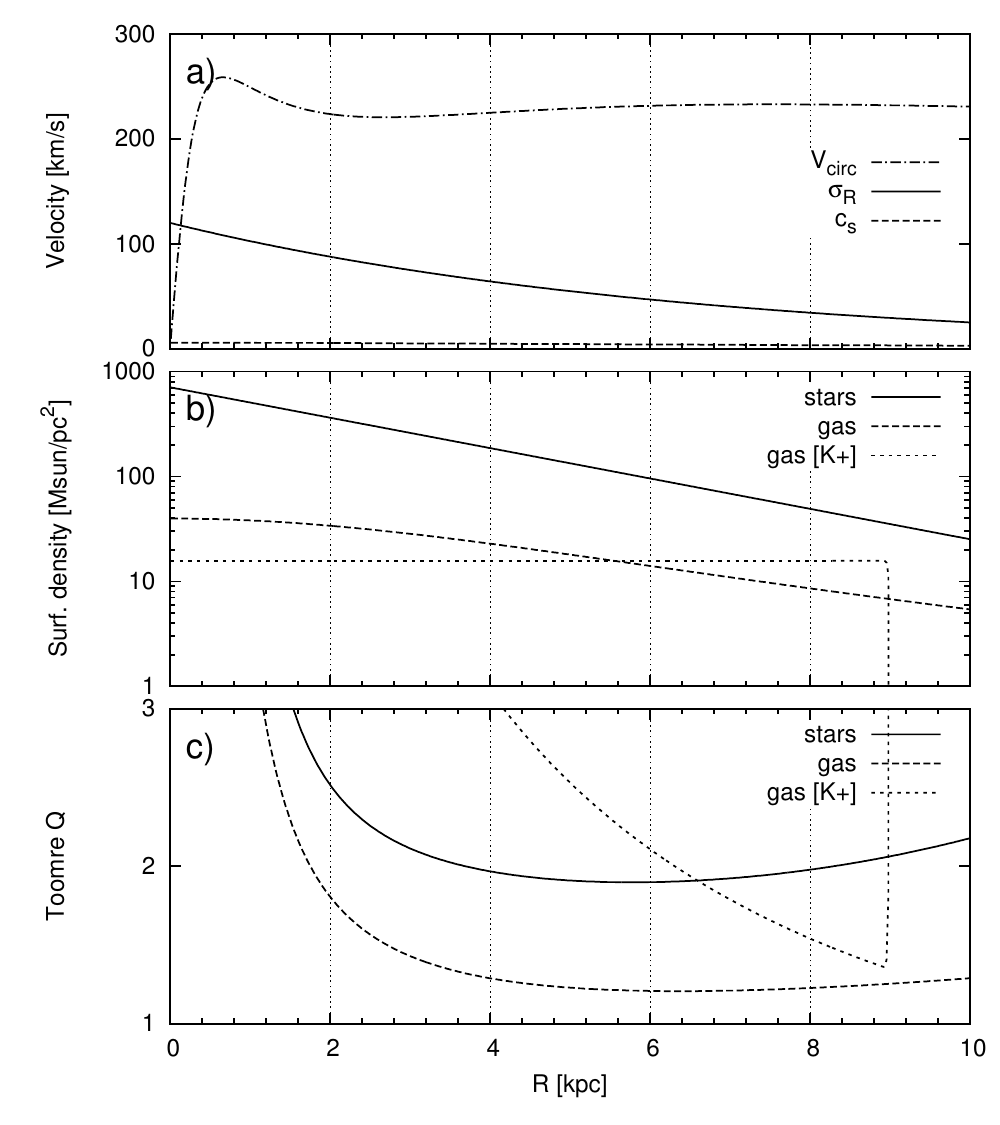}}
  \caption{The Galaxy model profiles: a) the rotation curve $V_\textrm{circ}(R)$, the radial dispersion of the velocities of the stars $\sigma_R(R)$ and the turbulent sound speed $c_\textrm{s}(R)$; b) the surface density of the stellar and gaseous discs; c) Toomre parameters(\ref{eq:t}) for stellar and gaseous discs. In panels b) and c) the short dashes show the corresponding profiles for the gaseous disc for the initial parameters adopted in K+.}
  \label{fig:vsqx}
\end{figure}

Fig.\,\ref{fig:vsqx}\,a shows profiles of the circular velocity $V_\textrm{circ}(R)$, the radial velocity dispersion of the stars $\sigma_R(R)$, and the turbulent sound speed $c_\textrm{s}(R)$. Fig.\,\ref{fig:vsqx}\,b shows the profiles of the surface density of the stellar and gaseous discs. 

Our rotation curve does not coincide with the rotation curve given in K+, and this can not be explained by the difference of the gas components. For example, a sharp drop in the rotation curve from K+ after the maximum at 1.2 kpc cannot be explained by the change of the gas distribution. Apparently, the curves given in K+ refer to the state of the system that has already left the initial non-equilibrium state.

This is also evidenced by the comparison of profiles of Toomre stability parameters for the stellar and gaseous discs presented in Fig.\,\ref{fig:vsqx}\,c:
\begin{equation}
   Q \equiv \frac{\kappa \sigma_R }{3.36 G \Sigma_\textrm{0}}\ ,\quad Q_\textrm{g} \equiv \frac{\kappa c_\textrm{s} }{\pi G \Sigma_\textrm{g0}}\ ,
   \label{eq:t}
\end{equation}
where $\kappa$ is the epicyclic frequency. Indeed, $Q(R)$ profile turned out to be close to the dependence given in K+ for the region $R>2$ kpc. However, $Q_\textrm{g}(R)$ profile for parameters adopted in K+ (short dotted line in Fig.\,\ref{fig:vsqx}\,c) differs from the behaviour of $Q_\textrm{g}(R)$ given in their Fig. 1, right panel. Since Toomre stability parameter is a measure of the dynamic temperature, $Q_\textrm{g}(R)$ profile in K+ reflects presence of rapid cooling, resulting in significantly lower sound speeds than 8 km/s.

Modelling of secular evolution and formation of spiral structures usually assumes stability of the stellar and gaseous discs with respect to axisymmetric perturbations. In non-interacting discs, it requires $Q, Q_\textrm{g}(R) \geq 1$. The interacting discs are more unstable, so the boundaries of stability in terms of the Toomre parameter shift towards higher values. However, if the discs are very different in mass and dynamical temperature, the correction terms turn out to be of the order $\epsilon \equiv \Sigma_\textrm{g0}/\Sigma_\textrm{0}$ (see Appendix). Thus, the presence of a stellar disc cannot be responsible for the rapid fragmentation of the gaseous disc.

In the model with fixed gas component, the stellar disc forms a bar (see Fig.\,5 of K+). A growth rate of bar formation depends on model parameters, such as a number of particles $N$ and a type of the halo. If the fixed halo potential (rigid halo) is substituted by a live halo, the growth rate increases by factor 2 or 3 \citep{PBJ16}. From the growth of $m = 2$ perturbation amplitude in K+, one can estimate the time of the exponential growth of the bar, which is $ T_\textrm{e} \approx 500$\,Myr. In a similar model with the live halo, \citet{P16_BA} received $T_\textrm{e}\approx 330$\,Myr. When gas simulation is turned on, a three-arm spiral in the stellar component appears during 170 ... 200\,Myr, which corresponds to a time scale $T_\textrm{e} \approx 40$\,Myr. Approximately the same period, 100 ... 150\,Myr, is required for formation of the three-arm spiral. This time interval is a typical dynamic time of the order of the rotation period at radii 4 ... 6 kpc.

\section{Gas disc fragmentation}

Stability analysis of the gaseous disc using matrix method by \citep{P17} shows absence of unstable global modes with growth rates $\omega_\textrm{I}> 1$\,Gyr$^{-1} $, or $T_\textrm {e}<1000$\,Myr. Therefore, the only dynamical mechanism that could lead to the rapid formation of clouds is the Jeans instability of the cold gaseous disc, i.e. one should anticipate the disc with $Q_\textrm{g} < 1$. A value of Toomre parameter that leads to fast fragmentation of the disc with $T_\textrm{e} \approx 40$\,Myr can be found from the linear stability theory. 

A preliminary estimate of this value can be made using WKB theory, assuming $m = 0$. From equation (A2) we find the most unstable wave number, $\hat k_\textrm{g} = 2/Q_\textrm{g}^2$, and corresponding growth rate
\begin{equation}
   \omega_\textrm{I} = \kappa(Q_\textrm{g}^{-2}-1)^{1/2}\ ,
\end{equation}
The loss of stability occurs at a radius of $R \approx 6.4$\,kpc corresponding to the minimum of $Q_\textrm{g}$, where $\kappa (R) = 53$\,Gyr$^{-1}$. Therefore, for the growth rate of $\omega_\textrm{I}=25$\,Gyr$^{-1}$, we obtain $Q_\textrm {g}\approx 0.9$.

\begin{figure}
  \centerline{\includegraphics[width = 85mm]{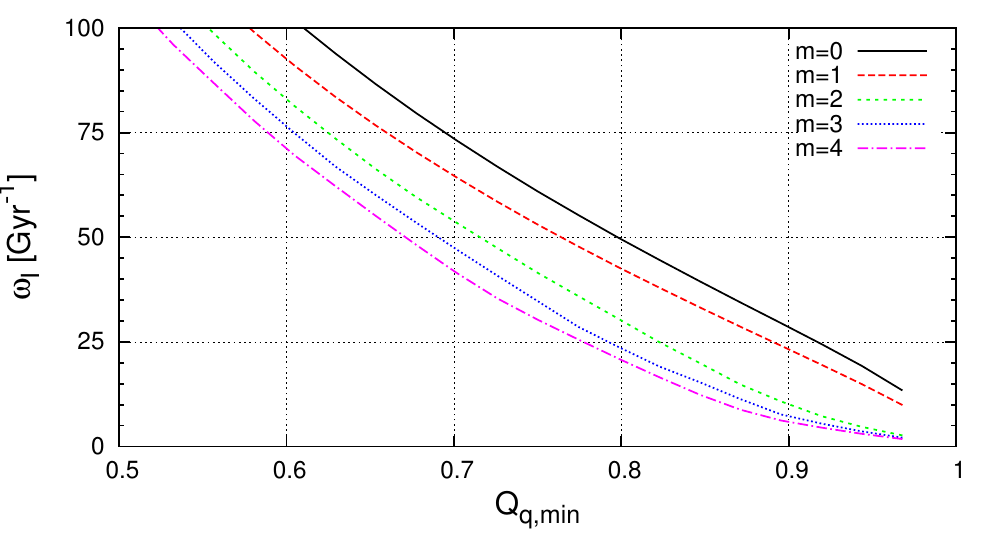}}
  \caption{The growth rate of the perturbations with the number of arms $ m = 0 ... 4 $ as a function of the minimum value of $ Q_\textrm{g}(R)$.}
  \label{fig:cw}
\end{figure}

Fig.\,\ref{fig:cw} shows the maximum growth rates for $m = 0 ... 4$ as the gas temperature varies, obtained with the matrix method for gaseous discs \citep{P17}. The unstable axisymmetric solution with $T_\textrm{e} = 40$\,Myr, or $\omega_\textrm{I} = 25$\,Gyr$^{-1}$, appears at $Q_\textrm{g,min} \approx 0.92 $, close to the estimate given above. The unstable three-arm spirals with this same $ T_\textrm{e} $ appears at $Q_ \textrm {g, min} \approx 0.79 $.

\begin{figure}
  \centerline{\includegraphics[width = 85mm]{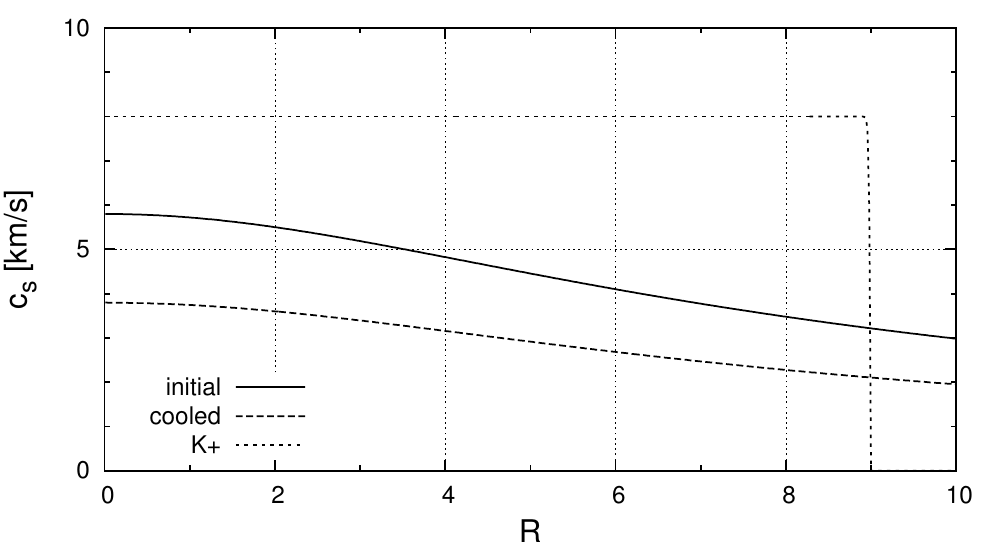}}
  \caption{Sound speed profiles $c_\textrm{s}(R) $, adopted in our model (solid line), in K+ at the initial moment (short dashed line) and corresponding to the cooled model with $ Q_\textrm{g, min} \approx 0.79 $ (dashed line).}
  \label{fig:cs}
\end{figure}

Fig.\,\ref{fig:cs} shows the sound speed profile corresponding to $ Q_\textrm{g, min} \approx 0.79 $ in the gaseous disc. For gas surface density $16$\,M$_\odot/$pc$^2$ at $R=5$\,kpc, $c_\textrm{s} \approx 4$\,km/s corresponds to $Q_\textrm {g} \approx 1.2$. Our calculations show that instability with the needed growth rate for harmonics $ m = 0 ... 3 $, requires gas cooling to $ c_\textrm{s} \approx 3$\,km/s.

\section{Formation of the stellar multi-arm spiral}

The cooling leads to gas fragmentation resulting in numerous molecular clouds. The latter affect stellar population of the galaxy. \citet{T81} reported on an effect of extraordinary rapid formation of two-arm spirals as a reaction to a quadrupole, time-limited perturbation of the stellar disc. In his model, the isolated disc was stable to the formation of spirals with any azimuthal numbers $ m $. The spirals appear due to the so-called {\it swing amplification} mechanism, which depends on two parameters, $ Q $ and $ X $,
\begin{equation}
	X \equiv \lambda_\theta/\lambda_\textrm{crit}\ ,
\end{equation}
$\lambda_\theta = 2\pi/k_\theta = 2\pi R/m$, $\lambda_\textrm{crit} = 4\pi^2 G \Sigma_0/\kappa^2$. As follows from the middle panel of Fig.7 in \citet{T81}, maximum amplification for $Q = 2$ corresponds to $X \approx 1.8 $, and the effect vanishes for $ X \gtrsim $ 3 \citep[see also][Fig.\,6.21]{BT}.

\begin{figure}
  \centerline{\includegraphics[width = 85mm]{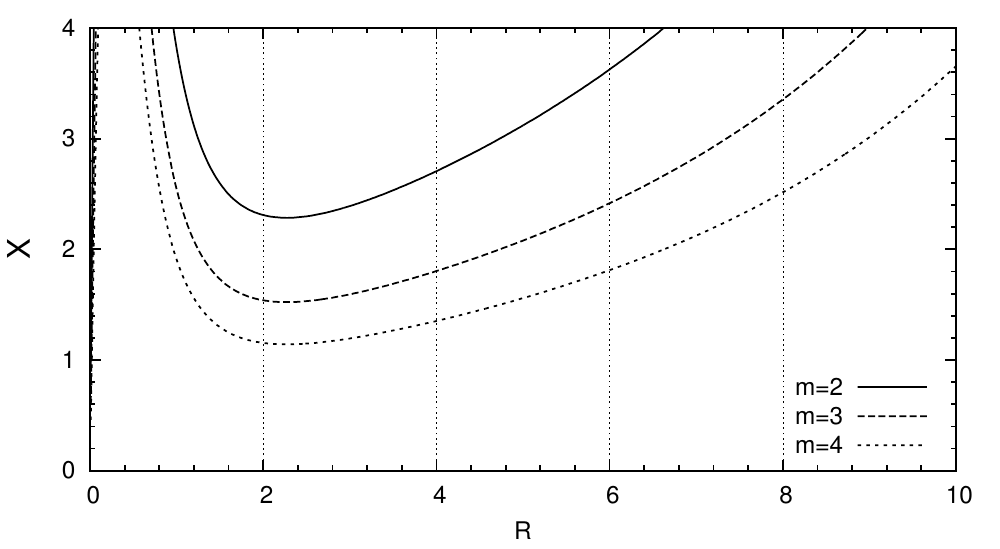}}
  \caption{The profiles $X(R)$ for $m=2,3,4$.}
  \label{fig:x}
\end{figure}

The profiles $ X (R) $ in Fig.\,\ref{fig:x} show that for $ m = 2 $, $ X (R) $ passes too high, so this mechanism can effectively support the two-arm spirals in the central region only. For $ m = 3 $, the value $ X = 1.8 $ occurs at 4.4 kpc, where a three-arm spiral is to be observed. It is precisely  the radius where such a three-arm spiral is observed in the numerical experiment by K+. Notice that a four-arm spiral on the periphery of the disc (at $ R> 7 $ kpc) is also possible.

\section{Conclusions}

Here we analyse the numerical experiments by \citet{KKK2016}, in which the appearance of a three-arm spiral in a thin stellar disc was demonstrated in the presence of a gaseous disc. Similar features of spirals are noted in a number of disc galaxies. We suggest that three-arm spirals can be explained by the rapid cooling of the gas component, which is accompanied by the appearance of molecular clouds. The latter, in turn, induce multi-arm spirals through the swing amplification mechanism.

As follows from the graphs of Toomre stability parameter $Q$ and $Q_\textrm{g} $, the initial value of the turbulent sound speed equal to 8 km/s, turn very quickly to 4 km/s at a radius $R=5$\,kpc. The needed cooling for the scenario described above is 3 km/s. We believe that this is quite realistic. In the future, we plan to test our hypothesis for multi-arm spiral formation using N-body simulations.

\section*{Acknowledgments}

This work was supported by the Sonderforschungsbereich SFB 881 ``The Milky Way System'' (subproject A6)
of the German Research Foundation (DFG), and by the Volkswagen Foundation under the Trilateral Partnerships grant No. 90411. The authors acknowledge financial support by the Russian Basic Research Foundation, grants 15-52-12387, 16-02-00649, and by Department of Physical Sciences of RAS, subprogram `Interstellar and intergalactic media: active and elongated objects'.

\section*{Appendix: Axisymmetric stability of interacting stellar and gaseous discs}

The isolated stellar and gaseous razor-thin discs in tightly-wound approximation are described by the dispersion relations (see, e.g., \citet{BT}):
\begin{align}
  &D_\textrm{s}(s,k) \equiv  s^2 - 1 + |\hat k_\textrm{s}|{\cal F} = 0\ , \tag{A1}\\
  &D_\textrm{g}(s,k) \equiv  s^2 - 1 + |\hat k_\textrm{g}| - \frac{Q_\textrm{g}^2 {\hat k}^2_\textrm{g}}{4} = 0\ ,  \tag{A2}
\end{align}
where 
\begin{equation*}
   s \equiv \frac{\omega_*}{\kappa}\ ,\quad \omega_* \equiv \omega-m\Omega\ ,\quad \hat k_\textrm{s} \equiv \frac{k}{k_\textrm{s, crit}}\ ,\quad \hat k_\textrm{g} \equiv \frac{k}{k_\textrm{g, crit}}\ ,
\end{equation*}
$k$ is a wavenumber, $k_\textrm{s, crit} = \kappa^2/(2\pi G \Sigma_{0})$,
$k_\textrm{g, crit} = \kappa^2/(2\pi G \Sigma_\textrm{g0})$, $c_\textrm{s}$ is the (turbulent) sound speed, 
${\cal F}$ is the reduction factor, which can be written explicitly for the Schwarzschild distribution function. Assuming $s=0$, we have:
\begin{equation}
   {\cal F}(0,\chi) = \frac 1\chi ( 1 - e^{-\chi} I_0(\chi) )\ ,\tag{A3}
\end{equation}
$\chi \equiv \sigma_R^2k^2/\kappa^2$, $\sigma_R$ is the radial velocity dispersion, $I_0$ is the modified Bessel function. 

In disc galaxies one typically has $\Sigma_\textrm{g0} \ll \Sigma_{0}$ and $ c_s \ll \sigma_R$, thus one can introduce two small parameters:
\begin{equation}
   \epsilon \equiv \Sigma_\textrm{g0} / \Sigma_{0}\ , \quad \delta \equiv c_\textrm{s} / \sigma_R\ .\tag{A4}
\end{equation}
Self-gravitation of the gaseous disc affects motion of stars and makes the stellar disc less stable. Loss of stability in isolated discs with respect to axisymmetric perturbations occurs at $Q = Q_\textrm{g}=1$, or $\hat k_\textrm{s}, \hat k_\textrm{g} \sim 1$. However, since $k_\textrm{s, crit}/k_\textrm{g, crit} = \epsilon$, the most vulnerable wave numbers in stars and gas are strongly separated, thus the resulting effect is negligible. Let's find the corresponding corrections to the Toomre parameters for interacting discs using a joint dispersion relation:
\begin{equation}
   D_\textrm{s}(s,k) D_\textrm{g}(s,k) = {\cal F} \hat k_\textrm{s} \hat k_\textrm{g}\ .\tag{A5}
\end{equation}
An analogous joint relation was obtained previously by \citet{R01} and \citet{RF13}, but analysed in a different manner.

In case of the stellar disc, we denote $\hat k = \hat k_\textrm{s}$, then $\hat k_\textrm{g} = \epsilon \hat k$, and the dispersion relation takes the form:
\begin{equation}
  s^2 =  1 - |\hat k| ({\cal F} + \epsilon) + {\cal O}(\epsilon^2)\ .\tag{A6}
\end{equation}
A small correction proportional to $\epsilon$ contributes to the critical value of the Toomre parameter in the presence of the gaseous disc:
\begin{equation}
  Q^\textrm{(i)} =  1 + 1.822 \epsilon + {\cal O}(\epsilon^2)\ .\tag{A7}
\end{equation}
Note that this expression doesn't depend on $Q_\textrm{g}$, since the critical wavelength is too large for the gas pressure to play any role.

For the gaseous disc, it is convenient to redefine $\hat k$ so that $\hat k = \hat k_\textrm{g}$. Thus, $\hat k_\textrm{s} = \epsilon^{-1} \hat k$, and the dispersion relation for the gaseous disc takes the form:
\begin{equation}
  s^2 =  1 - |\hat k| + \frac{Q_\textrm{g}^2 {\hat k}^2}{4} - 3.5 \frac \epsilon{Q^2} \left(\frac{1}{|\hat k|} 
  + \frac{Q^2_\textrm{g}}{4-Q^2_\textrm{g} |\hat k|}  \right) + {\cal O}(\epsilon^2)\ .\tag{A8}
\end{equation}
The critical wavelength at which the loss of stability occurs is given by the expression:
\begin{equation}
	|\hat k_*| = \frac{2}{Q^2_\textrm{g}}\left(1 - \frac{3.5\epsilon}{Q^2} \left[ \frac4{Q^4_\textrm{g}} - \frac{Q^4_\textrm{g}}4 \right]\right) + {\cal O}(\epsilon^2)\ ,\tag{A9}
\end{equation}
and the presence of the stellar disc shifts the critical parameter for the gaseous disc, $ Q_ \textrm{g}=1$, to:
\begin{equation}
  Q_ \textrm{g}^\textrm{(i)} =  1 + 1.75 Q^{-2} \epsilon + {\cal O}(\epsilon^2)\ .\tag{A10}
\end{equation}

\end{document}